\documentclass[11pt]{article}
\usepackage{moriond,epsfig}

\def\Journal#1#2#3#4{{#1} {\bf #2}, #3 (#4)}

\def\PRL{\em Phys. Rev. Lett.}

\def\MN{{\em Mon. Not. of the RAS}}
\def\ApJ{{\em Astrophys. Jour.}}
\def\ApJS{{\em Astrophys. Jour. Supp.}}
\def\AandA{{\em Astron. \& Astrophys.}}
\def\araa{{\em Ann. Rev. Astron. Astrophys.}}
\def\Nature{{\em Nature}}
\def\aas{{\em Am. Astron. Soc.}}

\def\be{\begin{equation}}
\def\ee{\end{equation}}
\def\bea{\begin{eqnarray}}
\def\eea{\end{eqnarray}}

\def\J{{\sl J} }
\def\H{{\sl H} }
\def\K{{\sl K} }

\def\k{{\sl K} }
\def\micron{${\rm{\mu}m}$ }

\def\wht{{\sl WHT} }
\def\int{{\sl INT} }
\def\iso{{\sl ISO} }
\def\chandra{{\sl Chandra} }
\def\scuba{{\sl SCUBA} }

\def\zs{$z_{\rm{spec}}$}
\def\in{$_{\rm{in}}$ }

\begin{document}
\vspace*{4cm}
\title{Multi-wavelength observations of serendipitous \chandra sources}

\author{P. Gandhi$^1$, C.S. Crawford$^1$, A.C. Fabian$^1$, R.J. Wilman$^{1,2}$, R.M. Johnstone$^1$, A.J. Barger$^{3,4,5,6}$ and L.L. Cowie$^{3,6}$}

\address{1. Institute of Astronomy, Madingley Road, Cambridge CB3 0HA, England\\
2. Leiden Observatory, P.O. Box 9513,  2300 RA Leiden, The Netherlands \\
3. Institute for Astronomy, 2680 Woodlawn Drive, Honolulu HI~96822, USA \\
4. Dept. of Astronomy, University of Wisconsin-Madison, 475 N Charter Street, Madison, WI~53706 USA \\
5. Hubble Fellow and Chandra Fellow at Large \\
6. Visiting Astronomer, W.M. Keck Observatory, jointly operated by the California Institute of Technology and the University of California }

\maketitle\abstracts{
We describe progress in a programme to study a sample of sources typical of those which contribute a large fraction of the hard X-ray background. The sources are selected from the fields of $\sim 10$ ks \chandra cluster observations with follow-up in the near-infrared and optical. The X-ray data indicate that many of these are powerful, obscured sources. Two hard objects which are lensed by the cluster are good candidates for X-ray type-2 quasars, with luminosities of $\sim 10^{44}-10^{45}$ erg s$^{-1}$ and obscuring column densities of $\sim 10^{23}$ cm$^{-2}$. We find that the sources are bright in the infra-red ($K\ {\rm typically}\ 17-18$). From Keck optical spectra and photometric redshifts, we find that the host galaxies are consistent with being early-type, massive hosts at redshifts ranging from 0.2 to 3, with $z_{\rm{median}}\approx 1.5$. The two obscured quasars also have mid-infrared detections and 850-\micron upper-limits implying that the surrounding dust is at warm-to-hot temperatures (T$\sim 1000$ K).  
}

\section{Introduction}

One of the first mysteries being probed by the \chandra X-ray Observatory~\cite{weisskopf} after its launch is the origin of the hard X-ray background. The superb sub-arcsecond resolution of the telescope coupled with good sensitivity (down to a $2-10$ keV flux $\sim 10^{-15}$ erg s$^{-1}$ cm$^{-2}$ in about 100 ks) has made the detailed study of the background source population possible. Many surveys~\cite{mushotzky}$^,$~\cite{hornschemeier}$^,$~\cite{tozzi} have revealed the multitude of sources which have evaded detection since the X-ray background was first discovered~\cite{giaconni}. These objects also have the correct, hard spectral slope to resolve the so-called \lq spectral-paradox' (see e.g. Fabian \& Barcons~\cite{acfxrb} for a discussion of the problem and a review of the observations before then). 

Efforts have since progressed to follow-up the source population at other wavelengths. We~\cite{c01a}$^,$~\cite{c01b} are in the midst of a programme to study the brighter (Flux$_{\rm{2-7\ {\rm keV}}} \sim 10^{-14}$ erg cm$^{-2}$ s$^{-1}$) of this source population. Our sources have been selected from the fields of a large sample of $\sim$ 10 ksec {\sl Chandra} cluster observations and lie at the break of the logN--logS relationship.~\cite{mushotzky} These objects are a subset of those found by Mushotzky et al., which contribute as much as $56-81\%$ of the $2-10$ keV background at around this flux level. We are targetting the sources with either faint, or undetected (down to $R\approx 24$) optical counterparts (hereafter referred to as optically-faint) in the near-infrared. We include, for comparison, some sources which have brighter optical counterparts. Synthesis models for the background predict that the spectral hardening is produced by obscuration surrounding an active-galactic nucleus~\cite{setti}$^,$~\cite{wilmanfabian}. We find that most objects are bright in the X-rays and the infra-red, yet show weak (or no) emission lines in the optical-near-infrared. Some of these objects would not have been identified as being AGN even in deep searches for optical emission lines. At least two of these are obscured quasars. 

\section{Observations}

\subsection{Infrared data}

\begin{figure}
\begin{center}
%\center{
%\rule{5cm}{0.2mm}\hfill\rule{5cm}{0.2mm}
%\vskip 2.5cm
%\rule{5cm}{0.2mm}\hfill\rule{5cm}{0.2mm}
\fbox{\psfig{figure=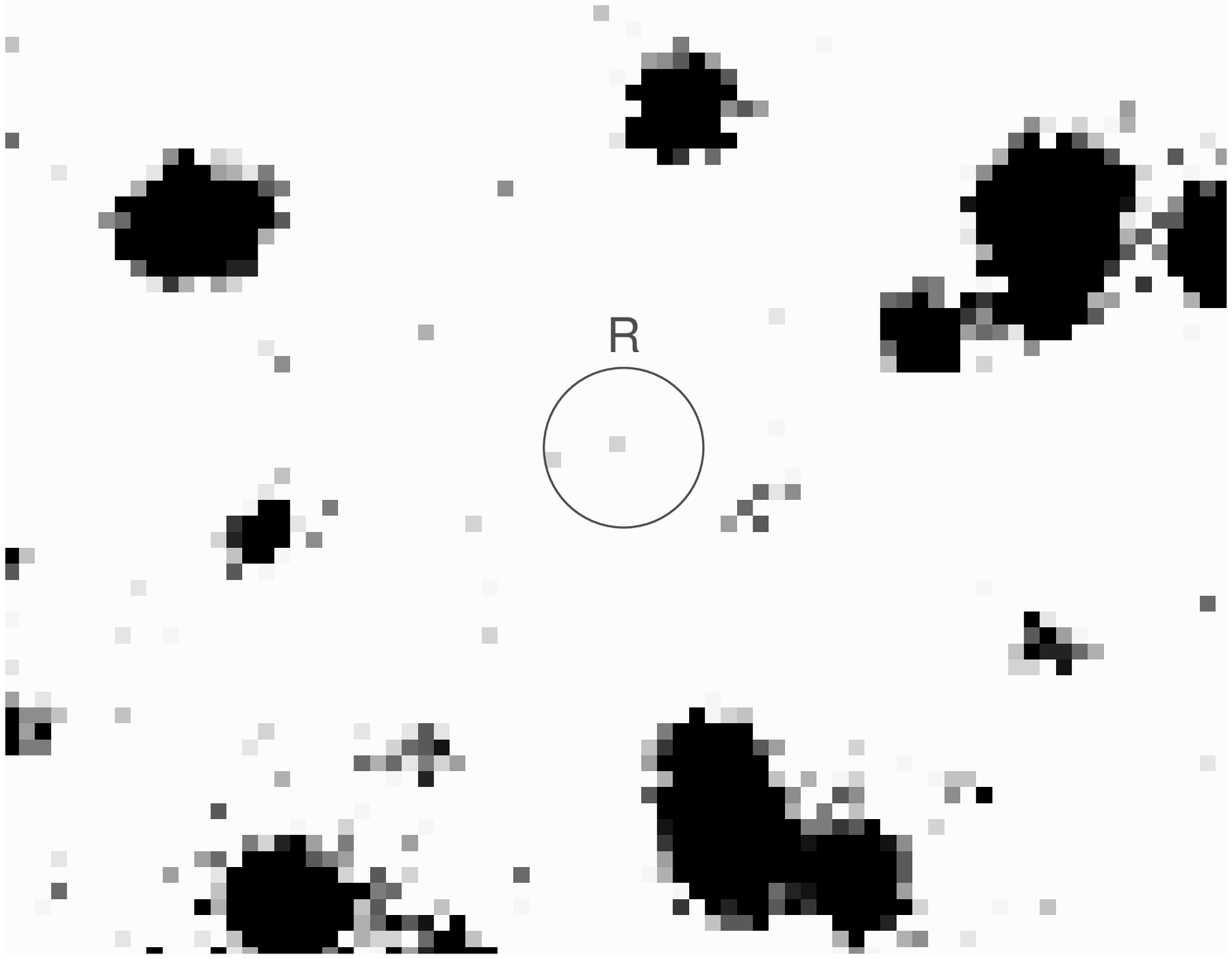,height=1.5in}}
\hskip .2cm
\fbox{\psfig{figure=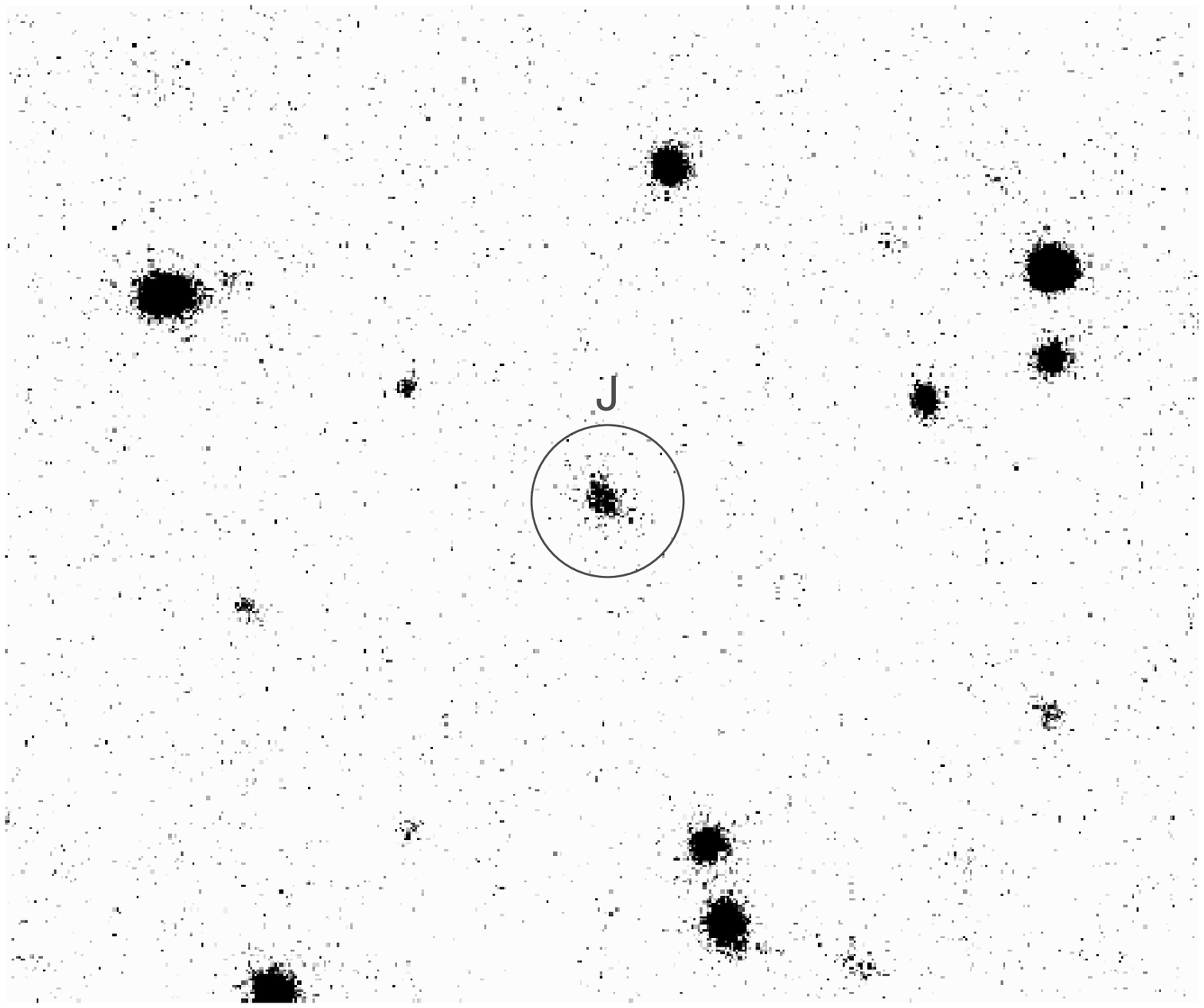,height=1.5in}}
\hskip .2cm
\fbox{\psfig{figure=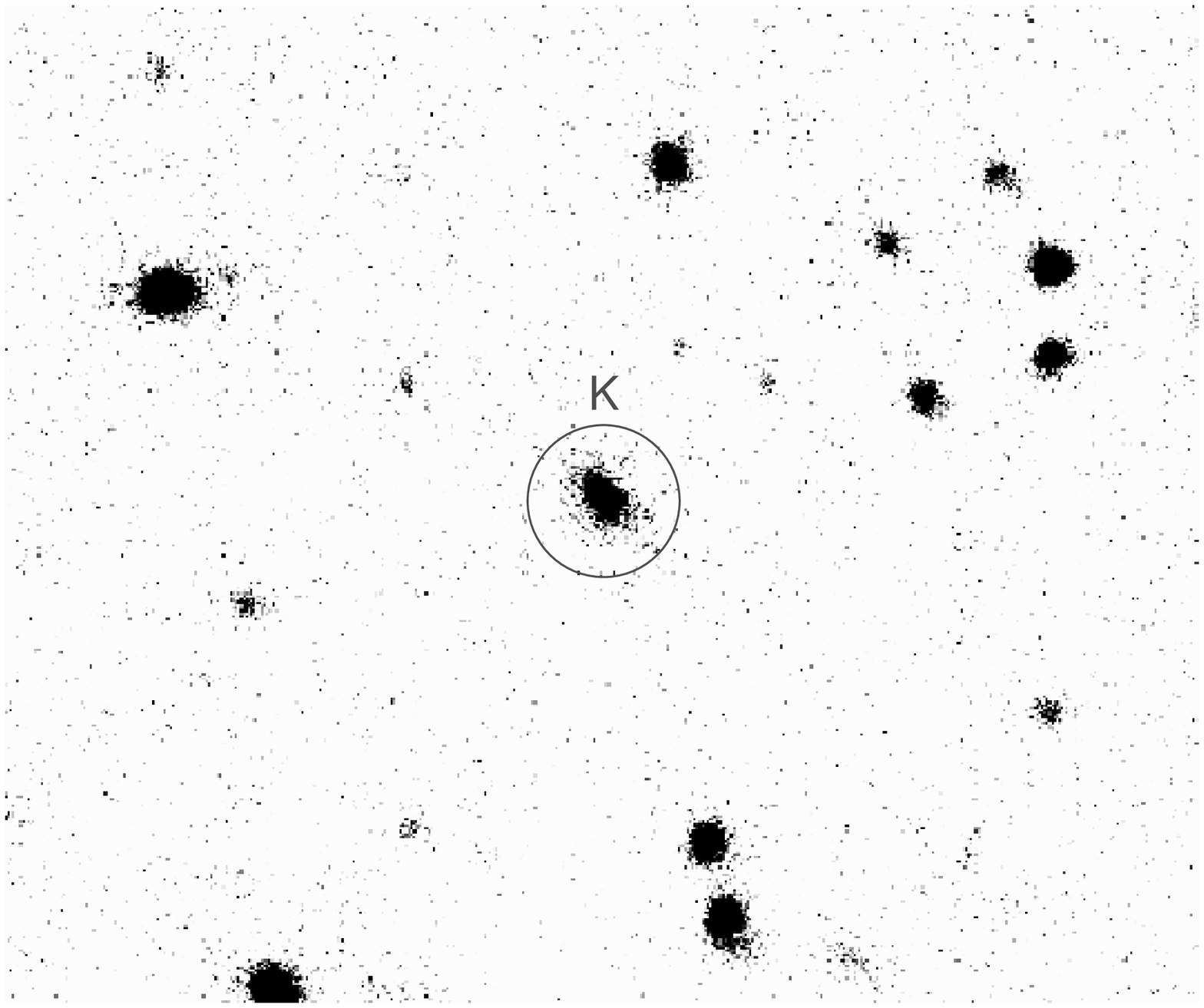,height=1.5in}}
\caption{Object A18 in the R, J and K filters (within the central circle), showing the optically-faint nature of this source. The box length is $\approx 35$ arcsec.
\label{fig:a18images}}
\end{center}
\end{figure}

The near-infrared observations were carried out at the United Kingdom Infrared Telescope (UKIRT). Full details of the observations can be referred to in Crawford et al.~\cite{c01a} (hereafter C01a) and Crawford et al.~\cite{c01b} (hereafter C01b). Spectroscopy with the cooled grating spectrograph (CGS4) revealed no strong emission features in any of six objects observed (see Figure 3 in C01a) in \J, \H or \K. If these harbour AGN within, we expect to have observed Paschen emission out to $z<1$, Balmer emission out to $z<3.5$ or MgII$\lambda$2798 at $3.2<z<6.8$. Imaging, on the other hand, revealed the objects to be fairly bright. Imaging was carried out with the $UFTI$ and $TUFTI$ arrays at UKIRT down to $K\approx 19$. To date, our catalogue of \K-band detections comprises almost 60 sources. \K$_{\rm{median}}\approx 17.5$. Figure~\ref{fig:a18images} shows, as an example, the morphology of source 18 (hereafter A18) in our sample of sources in the field of Abell 2390. This is also one of the reddest sources in our sample, with $R-K>7$.

\subsection{Optical data}

Available telescope archives were searched for the fields of all \chandra sources. These were typically the archives of the Isaac Newton Telescopes and the Canada-France-Hawaii Telescope.  Where such data was unavailable, the DSS and the APM were queried. In addition to this photometry, optical spectra were also obtained for fifteen sources in the Abell 2390 field with the Keck telescope (Figure~\ref{fig:keck}) and two sources in Perseus with the ISIS spectrograph on the \wht at La Palma.

\section{Photometric Redshifts}

Given the lack of significant emission features in the infrared spectra of these optically-faint objects, and the clear detections in imaging, we estimate redshifts by fitting the broadband optical-infrared spectrum to template spectra and searching for breaks or features in the continuum. A few caveats of this procedure should be kept in mind, however. Firstly, the fits depend, of course, on the template spectra available. Thus, it is essential to provide templates covering a wide region of parameter space and spectral types. Secondly, since the photometry is broadband, the effect of any narrow features will be averaged out. In fact, the photometric-redshift fitting procedure should work best for objects with no obvious strong central active-nuclei (AGN), which may be variable and typically show a power-law spectrum with no obvious continuum breaks. It is also difficult to estimate the contributions of the host galaxy and of the AGN to the total continuum, thus making it difficult to provide approriate templates. Finally, there is also the well-known degeneracy between reddened galaxies which are red due to intrinsic obscuring matter and those that are red due to advanced age.

\begin{figure}
\begin{center}
%\rule{5cm}{0.2mm}\hfill\rule{5cm}{0.2mm}
%\vskip 2.5cm
%\rule{5cm}{0.2mm}\hfill\rule{5cm}{0.2mm}
\psfig{figure=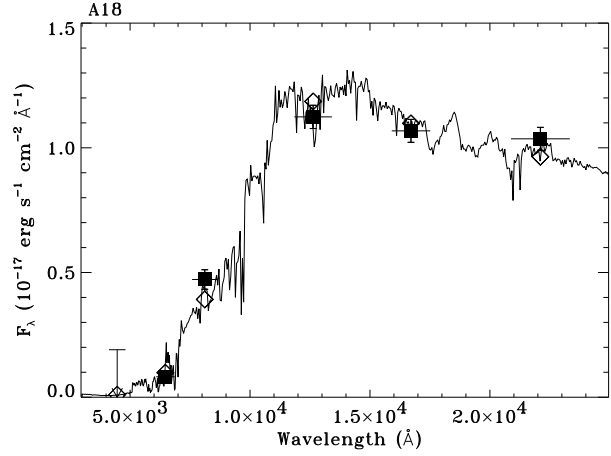,height=2.5in}
\caption{The broadband spectral energy distribution of object A18. The arrow denotes the B-band 3-$\sigma$ upper-limit and detections are shown for the R, I, J, H and K filters. The x - error bars denote the bandwidths, while the unfilled diamonds represent the integrated flux of the template SED through the individual filters. The best-fit SED is a Bruzual \& Charlot early-type host spectral model at $z_{\rm phot}=1.45$.
\label{fig:a18}}
\end{center}
\end{figure}

\begin{figure}
\begin{center}
%\rule{5cm}{0.2mm}\hfill\rule{5cm}{0.2mm}
%\vskip 2.5cm
%\rule{5cm}{0.2mm}\hfill\rule{5cm}{0.2mm}
\psfig{figure=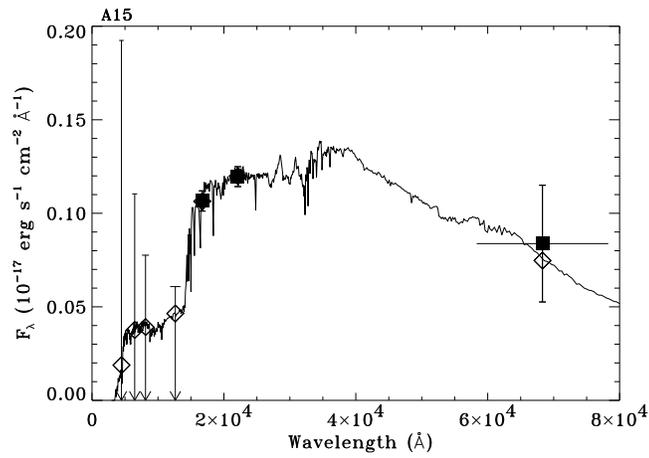,height=2.5in}
\caption{Best-fit Bruzual \& Charlot elliptical galaxy spectral model for object A15 at $z_{\rm{phot}}=2.8$. Upper-limits are shown for B, R, I and J, while detections are shown as solid squares for H, K and the 6.7-\micron \iso filter.
\label{fig:a15}}
\end{center}
\end{figure}

We used the publicly-available code HYPERZ~\cite{bmp} which fits any number of template spectra to broadband photometry in specified filters. The redshift, age and intrinsic reddening of the templates is varied in order to obtain the minimum-$\chi^2$ solution. The template spectra used comprised Bruzual \& Charlot~\cite{bc} synthetic spectra, Coleman-Weedman-Wu~\cite{cww} empirical spectra, a high signal-to-noise average QSO spectrum~\cite{francis} and host-galaxy-models with various contributions of reddened power-laws (see C01b). An example of a HYPERZ fit is shown in Figure~\ref{fig:a18} for object A18 (the object in Figure~\ref{fig:a18images}) from the Abell 2390 cluster field. HYPERZ fits an evolved early-type galaxy model at $z_{\rm phot}=1.45$, in very good agreement with the spectroscopic redshift(\zs)$=1.467$ obtained by Cowie et al.~\cite{cowie} for this very red object. Figure~\ref{fig:a15} shows another red source (A15) in the field of the A2390 which has a maximum-likelihood solution of $z_{\rm phot}=2.78$, this time in good agreement with a photometric redshift obtained by Cowie et al. This galaxy shows evidence for intrinsic reddening of about A$_{\rm V}=1.8$ magnitudes.

\begin{figure}
\begin{center}
%\rule{5cm}{0.2mm}\hfill\rule{5cm}{0.2mm}
%\vskip 2.5cm
%\rule{5cm}{0.2mm}\hfill\rule{5cm}{0.2mm}
\psfig{figure=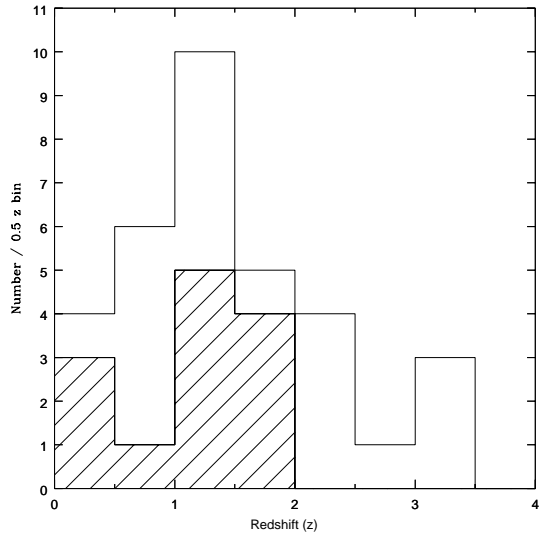,height=3.in}
\caption{The distribution of redshifts obtained for a sample of sources from the A2390, A2199, A1795, A1835, Perseus and IRAS09104+4109 fields. The shaded region represents the sources with a spectroscopic redshift measurement, the rest being photometric estimates.
\label{fig:zhist}}
\end{center}
\end{figure}

\begin{figure}
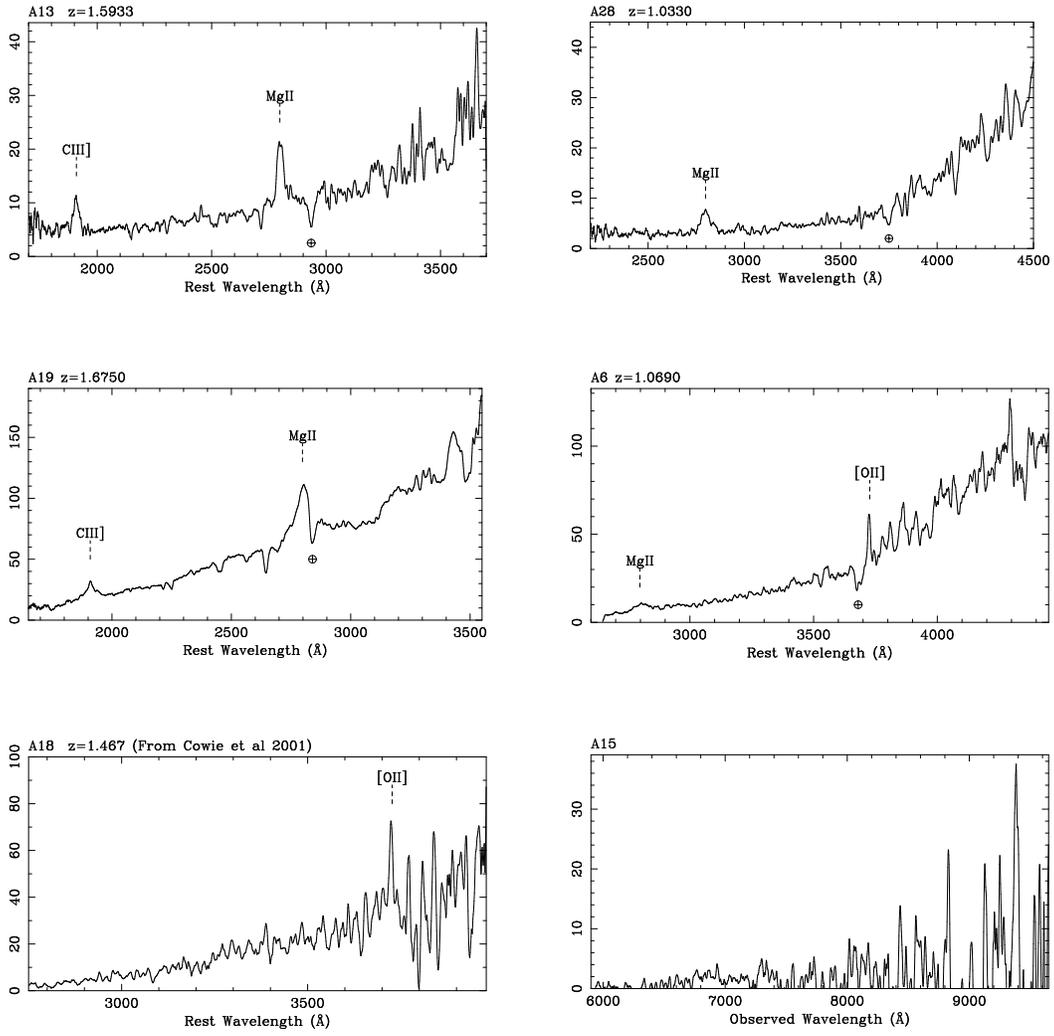

\begin{center}
%\center{
%\rule{5cm}{0.2mm}\hfill\rule{5cm}{0.2mm}
%\vskip 2.5cm
%\rule{5cm}{0.2mm}\hfill\rule{5cm}{0.2mm}
\rotatebox{270}{\psfig{figure=kecka13.ps,height=2.5in}}
\hskip 1cm
\rotatebox{270}{\psfig{figure=kecka28.ps,height=2.5in}}
\hfill
\vskip 1.cm
\rotatebox{270}{\psfig{figure=kecka19.ps,height=2.5in}}
\hskip 1cm
\rotatebox{270}{\psfig{figure=kecka6.ps,height=2.5in}}
\hfill
\vskip 1.cm
\rotatebox{270}{\psfig{figure=kecka18.ps,height=2.5in}}
\hskip 1cm
\rotatebox{270}{\psfig{figure=kecka15.ps,height=2.5in}}
\hfill
\caption{Keck Spectra for a sample of sources at $z>1$. The softer of these sources all showed emission lines. The last source (A15) is shown in observed-wavelength space since we could only obtain a photometric estimate for this.
\label{fig:keck}}
\end{center}
\end{figure}

Despite the caveats mentioned, we obtained good agreement between the photometric redshift estimates and spectroscopic measurements where both were available.This is primarily because many of these objects show large obscuration toward their centres, which implies that the optical/near-infrared emission is dominated by the host galaxy and not any AGN present therein. The objects for which there was a mismatch between the two redshifts are also the sources which show signatures of being relatively unobscured AGN and are discussed further in C01b. The redshift distribution extends out to $z\approx 3.5$, with the median lying at about 1.4 (Figure~\ref{fig:zhist}). We caution that some of the highest redshift estimates are based on sparse photometry in the optical/near-infrared. We hope to be able to include more data on these objects in the future to check the robustness of these estimates.

\section{The nature of the sources}

\subsection{Evidence from X-rays}

\begin{figure}
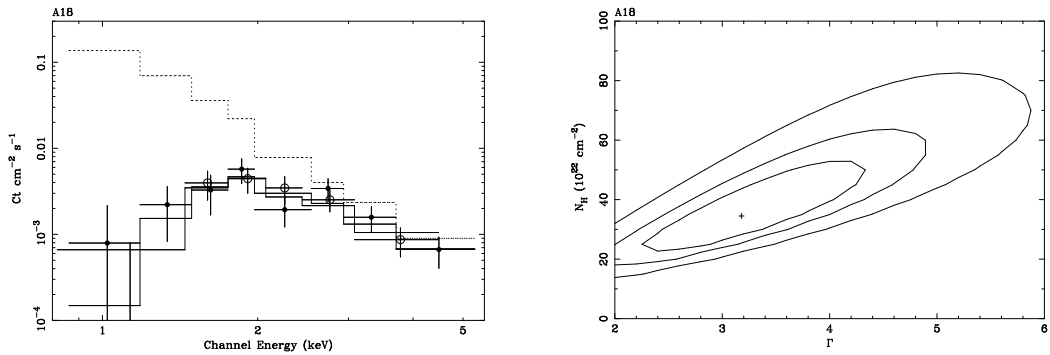

\begin{center}
%\center{
%\rule{5cm}{0.2mm}\hfill\rule{5cm}{0.2mm}
%\vskip 2.5cm
%\rule{5cm}{0.2mm}\hfill\rule{5cm}{0.2mm}
\rotatebox{270}{\psfig{figure=A18xray_spec.ps,height=2.5in}}
\hskip 1.cm
\rotatebox{270}{\psfig{figure=A18xray_conf.ps,height=2.5in}}
\caption{X-ray data for A18 (circles from two observations) with the best-fit absorbed (intrinsic + Galactic) power-law model (solid line) shown in the plot on the left. A spectrum with only Galactic absorption (dotted line) is clearly unacceptable. This is also shown on the right, with the 1-, 2- and 3-sigma contours indicating the need for intrinsic obscuration for all reasonable values of the power-law index $\Gamma$.
\label{fig:a18xray}}
\end{center}
\end{figure}

The near-infrared colours of our sample are different from those of stellar objects. Is there evidence for these being obscured AGN, as predicted by synthesis models? The strongest such evidence probably comes from the X-ray observations. Figure~\ref{fig:a18xray} shows the \chandra ACIS-S spectrum of source A18 fitted to an intrinsically-absorbed power-law. The best-fit model is obscured under a column-density of N$_{\rm{H}}=2\times 10^{23}$ atoms cm$^{-2}$, with a high $2-10$ keV luminosity of 10$^{45}$ erg s$^{-1}$ (at \zs=1.467; H$_0=50$; q$_0=0.5$). Such a luminous, point X-ray source could only be a quasar. We refer to it as an X-ray type-2 quasar. It is the large amount of (Compton-thin) obscuration here which makes this a type-2 object, rather than the existence of narrow emission lines as in the corresponding family of Seyferts. A15 is another highly obscured source, with a column of $2\times 10^{23}$ cm$^{-2}$ and a luminosity of $2\times 10^{44}$ erg s$^{-1}$. It should be noted that both these objects lie closest to the central cD galaxy of the cluster and are magnified by a factor of 2.1 and 7.8 respectively, thus boosting the counts that we observe. The luminosities stated have been corrected for this factor. There are models which propose that the X-ray background emission could be from the most active phase in the life-cycle of a galactic nucleus (e.g. Fabian~\cite{fabian}). This is the period of growth of the nucleus through radiatively-efficient accretion of surrounding matter, naturally giving rise to very powerful, very obscured objects.

Most of the other sources for which spectral fits to the X-ray data could be performed indicated column densities $\sim 10^{21}-10^{22}$ cm$^{-2}$, and many show AGN-{\it like} emission lines in the optical. A sample of Keck spectra for sources at $z>1$ is shown in Figure~\ref{fig:keck}. All the objects at $z>1$ which show readily-detectable emission lines have soft X-ray spectra, the harder sources possessing higher column densities where these could be measured. It is this higher obscuring column which could be responsible for the depletion of the lines in the harder sources. Another possible reason that these objects (and many hard X-ray background source in general) do not show strong AGN emission lines may be the result of dust within narrow-line region clouds.~\cite{netzer} As we will show in the next section, some of these sources may have a large amount of dusty gas surrounding them.

A large fraction of the sources also showed definite variability (by as much as a factor of 4) between two X-ray observations about one year apart -- another strong indication that they harbour AGN.

\subsection{Evidence from the mid-infrared}

There is another clue which gives credence to the obscured AGN hypothesis. If these objects have a large amount of matter surrounding them, any absorbed energy is probably re-emitted at longer wavelengths after reprocessing. The \scuba sub-mm camera on the James Clerk Maxwell Telescope has played a significant role in unveiling obscured star-formation in the Universe, especially at high redshift~\cite{blain}. However, a distinctly weak correlation has been found between \chandra detections and their \scuba counterparts~\cite{f00}$^,$~\cite{bargeretal}, along with definite detections in the mid-infrared \iso 6.7- and 15-\micron bands~\cite{lemonon}$^,$~\cite{aussel}. Could this imply that the temperature of the reprocessing material is hotter, i.e. more characteristic of mid-infrared wavelengths, rather than sub-mm? In order to check this hypothesis, we performed a radiative-transfer-through-dust calculation~\cite{wfg} using the public code DUSTY~\cite{ine}. Given a primary optical/UV continuum (modelling the big-blue-bump) of the central AGN, as well as the geometry, compostion, size and density of the surrounding dust particles, DUSTY calculates radiative transfer spectra for the variables T\in at r\in (the temperature at the inner-most radius of the dust distribution), $\tau$ (the optical depth) and r$_{\rm{out}}$/r\in (the scale of the distribution). DUSTY assumes a spherical geometry for the surrounding dust. This is probably a good assumption for the more powerful, high column-density sources, since it is possible~\cite{fi} that a very large fraction of the energy produced by accretion (as much as 85\%) is obscured.

\begin{figure}
\begin{center}
%\center{
%\rule{5cm}{0.2mm}\hfill\rule{5cm}{0.2mm}
%\vskip 2.5cm
%\rule{5cm}{0.2mm}\hfill\rule{5cm}{0.2mm}
\rotatebox{90}{\psfig{figure=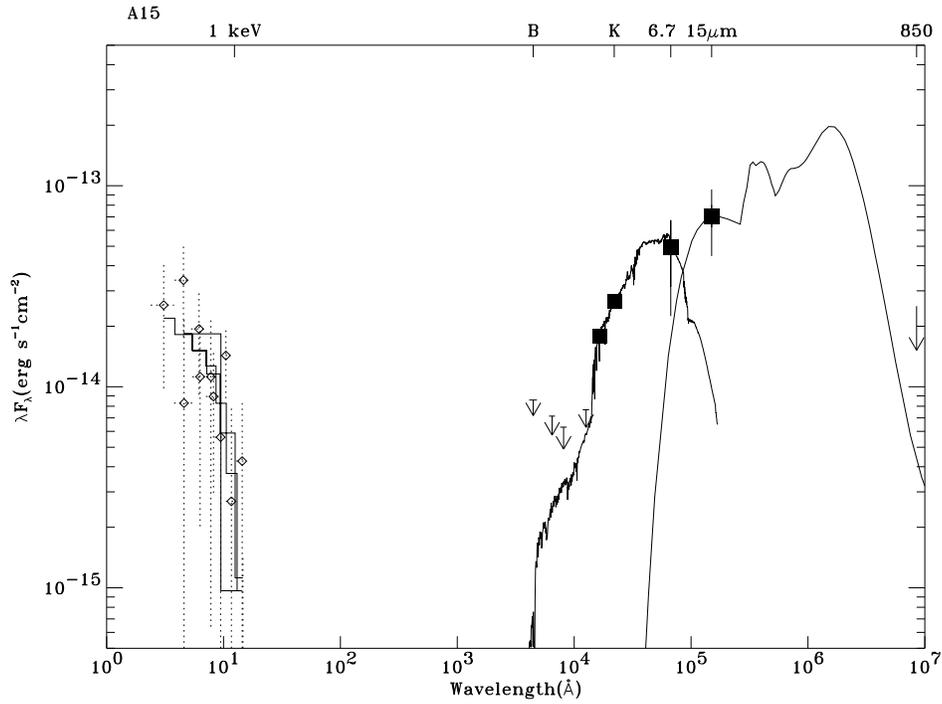,height=5.in}}
\caption{The total spectral energy distribution
for A15. The X-ray data are marked by open diamond markers with 
error bars. Our optical and near-infrared values and limits are marked
as solid squares. The best-fit absorbed power-law model to the X-ray spectrum is shown as a solid line, and the other solid
lines show the fits obtained in HYPERZ (optical to near-infrared) and
DUSTY (normalized to the 15-\micron flux). The {\sl SCUBA} upper-limit is shown as the arrow at 850-\micron.
\label{fig:a15dusty}}
\end{center}
\end{figure}

We fit the models against mid-infrared detections of sources in the deepest \iso field observed -- that of A2390~\cite{altieri}. Figure~\ref{fig:a15dusty} shows the total spectral energy distribution for source A15. The solid line over the mid- to far-infrared region represents the reprocessed spectrum as computed by DUSTY, along with the best-fit HYPERZ SED in the optical-near-infrared (solid line in the optical/near-infrared). The DUSTY spectrum has been redshifted according to the HYPERZ photometric redshift and normalised to the 15-\micron \iso flux~\cite{lemonon}. The arrow in the sub-mm shows the upper-limit to the 850-\micron \scuba flux. The best models all had inner-dust temperatures of $>$1000 K. Reducing the temperature significantly begins to cut into the 850-$\mu$m flux and also miss the 6.7-$\mu$m flux completely. Though some models of lower temperatures were also acceptable, many of these underestimate the 850-\micron flux by only a factor of a few. Given that most of the \chandra sources are undetected at sub-millimeter wavelengths, the 850-\micron flux is likely to lie well below the \scuba limit (see Wilman, Fabian \& Gandhi~\cite{wfg} for more details). Observations from the new generation of mid-infrared telescopes (such as at 70-\micron with {\em SIRTF}~\cite{sirtf}) should narrow the temperature range further. 

The DUSTY model in Figure~\ref{fig:a15dusty} has T$_{\rm in}=1500$ K and $\tau$($0.3\mu$m)$=40$, with the intrinsic optical/UV luminosity = $3\times 10^{45}$ erg s$^{-1}$ and the dust extending from 0.1 parsec out to 50 parsec from the central source. The inferred intrinsic optical-to-Xray spectral index $\alpha_{\rm{ox}}$ between 2 keV and 2500 angstroms is 1.3 (defined such that $F_{\nu}\propto \nu^{-\alpha_{\rm{ox}}}$), in good agreement with the intrinsic spectral index found by Elvis et al.~\cite{elvis} by studying a large sample of bright, unobscured {\em EINSTEIN} quasars. Thus, we can say that the (sub-mm) mid-infrared (non-) detections are consistent with a model for an obscured AGN with energy being reprocessed from the optical-UV to the mid- to far- infrared, and reprocessing dust at warm-to-hot temperatures extending in to within a parsec of the central AGN.

\subsection{The host galaxy} 

\begin{figure}
\begin{center}
%\center{
%\rule{5cm}{0.2mm}\hfill\rule{5cm}{0.2mm}
%\vskip 2.5cm
%\rule{5cm}{0.2mm}\hfill\rule{5cm}{0.2mm}
\rotatebox{90}{\psfig{figure=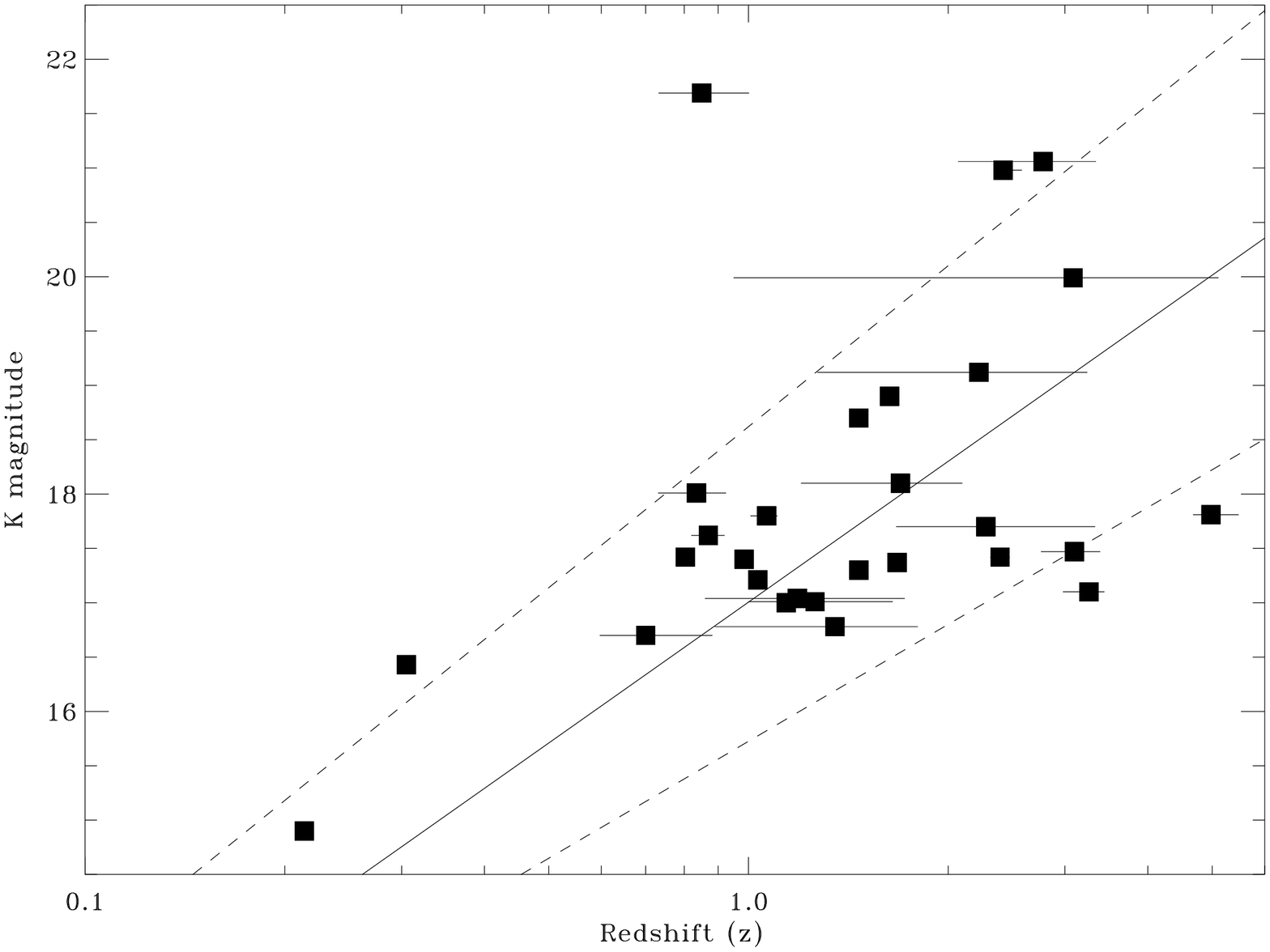,height=3.5in}}
\caption{K-magnitude vs. Redshift for a sample of our objects from the A2390, A2199, A1795, A1835, Perseus and IRAS09104+4109 fields. The 90\% confidence interval is shown for objects with only a photometric estimate. The solid line shows the K-z relation for massive radio ellipticals, and the dotted lines indicate the scatter of massive ellipticals about this line.
\label{fig:kz}}
\end{center}
\end{figure}

Most of the best-fit HYPERZ spectra to the optical/near-infrared photometry are early-type host galaxy models with exponentially-decaying star-formation timescales of about a gigayear. There is a well-known relation between the \k magnitude of the host-galaxies of quasars and their redshift. Figure~\ref{fig:kz} shows the $K-z$ relation for those sources in our \chandra fields with either a photometric or a spectroscopic redshift. The solid line is a rough-fit to the $K-z$ relation which has been observed for massive elliptical galaxies (see Eales et al.~\cite{eales}), while the dashed lines are indicative of the scatter in the Eales et al. sample. Most of our objects are consistent with being bright, and presumably massive, galaxies - in fact, slightly brighter than an L$_*$ galaxy.

\section{Conclusions}

We are compiling a sample of serendipitous \chandra sources with X-ray-brightness close to the break in the hard X-ray source counts ($S_{2-7\ {\rm kev}}\sim 10^{-14}$ erg cm$^{-2}$ s$^{-1}$) . Most show evidence of Compton-thin AGN, with a few having N$_{\rm{H}}\sim 10^{23}$ cm$^{-2}$ and L$_{\rm{intrinsic}}\sim 10^{45}$ erg s$^{-1}$. These very powerful objects are optically-faint ($R\ge 24$) and typically lie in massive, red, early-type host galaxies. The sources lie at all redshifts from $\sim 0.2-3$, with a distribution peaking around z$_{\rm{median}}\approx 1.5$. Heavy obscuration to the nucleus implies that the optical/near-infrared emission is dominated by the host galaxy and not the AGN, thus making the observation of strong emission features difficult, though weaker emission lines are seen in the less obscured sources -- many of them indicating the possible presence of an AGN. The non-detections of \chandra sources with \scuba and the presence of mid-infrared \iso counterparts indicates warm-to-hot dust, being heated by an AGN, with dust clouds extending to within $\sim$ 1 parsec of the centre.

Thus, through revealing the X-ray background, we might be viewing the period of rapid and violent growth of massive black holes within large galaxies.

The number of objects in the Universe which would be classified as AGN has increased significantly within the first year of the launch of {\it Chandra}. There are very exciting prospects for the future X-ray missions which will probe the peak of the X-ray background at energies of around 30 keV.

\section*{Acknowledgments}

PG thanks for the Sir Isaac Newton Trust and the Cambridge University Overseas Research Scheme for financial support. CSC and ACF thank the Royal Society. RJW and RMJ thank PPARC. Support for AJB was provided by NASA through the Hubble Fellowship
grant HF-01117.01-A awarded by the Space Telescope Science
Institute, which is operated by the Association of Universities
for Research in Astronomy, Inc., for NASA under contract
NAS 5-26555.  AJB and LLC acknowledge support from NSF through
grants AST-0084847 and AST-0084816, respectively.

\end{document}